\documentclass[twocolumn,prb]{revtex4-2}
\usepackage{graphicx,amssymb,soul,color,amsmath,bm}
\usepackage{epstopdf,hyperref}
\usepackage{braket,dsfont,nicefrac}
\usepackage[mathcal]{euscript}
\usepackage[normalem]{ulem}
\usepackage[utf8]{inputenc}

\hypersetup{colorlinks=true,citecolor=blue,linkcolor=magenta}

\sethlcolor{yellow}
\allowdisplaybreaks

\begin{document}

\title{Neural network representation for minimally entangled typical thermal states}

\author{Douglas Hendry}
\affiliation{Department of Physics, Northeastern University, Boston, Massachusetts 02115, USA}
\author{Hongwei Chen}
\affiliation{Department of Physics, Northeastern University, Boston, Massachusetts 02115, USA}
\author{Adrian Feiguin}
\affiliation{Department of Physics, Northeastern University, Boston, Massachusetts 02115, USA}

\begin{abstract}
Minimally entangled typical thermal states (METTS) are a construction that allows one to to solve for the imaginary time evolution of quantum many body systems. By using wave functions that are weakly entangled, one can take advantage of efficient representations in the form of matrix product states. We generalize these ideas to arbitrary variational wave functions and we focus, as illustration, on the particular case of restricted Boltzmann machines. The imaginary time evolution is carried out using stochastic reconfiguration (natural gradient descent), combined with Monte Carlo sampling. Since the time evolution takes place on the tangent space, deviations between the actual path in the Hilbert space and the trajectory on the variational manifold can be important, depending on the internal structure and expressivity of the variational states. We show how these differences translate into a rescaled temperature and demonstrate the application of the method to quantum spin systems in one and two spatial dimensions.  
\end{abstract}
\maketitle

\section{Introduction}

Quantum Monte Carlo is a powerful method that can be applied to problems with hundreds of degrees of freedom in arbitrary dimensions. However, while in principle can be considered an unbiased numerical technique, in many cases it suffers from some pathological drawbacks. Among these, there is the infamous sign problem, that appears in fermionic and frustrated systems, when the complicated nodal structure of the wave function does not guarantee a well defined positive transition probability. Other situations arise when the Monte Carlo updates necessary to make the simulation ergodic are complicated or numerically costly, or when the system is close to a phase transition and global updates are required to fight critical slowing down. 

Alternatives to QMC that can overcome such drawbacks are not many, and also suffer from limitations. Useful practical approaches that rely on exact diagonalization are limited to small system sizes \cite{drabold93,jaklic94,aichhorn03,long03,weise06,avron11,sugiura13,hanebaum14,hyuga14,roosta-khorasani15,saibaba17,sugiura17,okamoto18,schnack20,Weinberg2021}. In quasi one-dimensional systems, a family of methods based on the density matrix renormalization group can essentially provide numerically exact results for models with frustration \cite{Nishino1995, Bursill1996, Wang1997, Maisinger1998, Nishino1999, verstraete2004renormalization, Feiguin2005a, Huang2012, Bruognolo2015, Chen2018} and recent advances have put quasi two-dimensional frustrated lattice problems within reach \cite{Chen2019}. 
At this point, it is important to point out that these approaches rely, one way or another, on representations of the transfer matrix or the thermal density matrix of the quantum many-body problem in the form of matrix product states (MPS) or matrix product operators and, as a consequence, they are limited by the entanglement growth in the system as temperature is lowered, implicitly imposing a numerical barrier that is hard to overcome. Recent proposals using entanglement purification with neural networks provide an interesting alternate route \cite{nomura2021purifying}.

In a thermal state, the expectation value of observables is identical to the value in the canonical ensemble at some temperature $T$. This idea lies at the foundation of the statistical mechanics and the canonical to micro-canonical correspondence, relating the thermodynamic behavior of systems at temperature $T$ to the microstates of a system at some energy $E(T)$. A generic chaotic closed system out of equilibrium is expected to relax to a thermal state after some time. This problem does no require a thermal bath, and in the context of the microcanonical ensemble and energy is conserved.  However, in order for this to actually occur, certain conditions need to be satisfied: The expectations values of observables within an energy window around $E(T)$ need to vary smoothly, or rather, to be very ``similar''. This is the premise behind the eigenstate thermalization hypothesis (ETH)\cite{Srednicki1994,Deutsch2018}, and the idea of ``typicality''. According to this, a thermal state can be represented accurately by a typical pure state in the microcanonical ensemble. This can be exploited to carry out finite temperature calculations with pure states, which is the foundation behind ``minimally entangled typical thermal states'' (METTS)\cite{Stoudenmire2010,White2009} and ``canonical thermal pure quantum states''(CTPQS)\cite{Sugiura2013,Hyuga2014,Sugiura2017}. In a nutshell, the recipe is very simple: start from a random state, such a linear combination of basis states with random coefficients $|\psi_0\rangle$, and evolve it with the operator $|\psi(\beta)\rangle=\exp{(-\beta H/2)}|\psi_0\rangle$ ($\beta$ as customary represents the inverse temperature $T$). Observables and hence obtained as $\langle \hat{A} \rangle _T = \langle \psi(\beta)|\hat{A}|\psi(\beta)\rangle/\langle \psi(\beta)|\psi(\beta)\rangle$. In the case of METTS, the initial random states are product states, {\it e.g.} quantum spins pointing in random directions on the Bloch sphere. 
In a nutshell, the algorithm is identical to projector Monte Carlo\cite{Blankenbecler1983, Trivedi1989}, but with the initial state being evolved using a numerically exact method.
On the other hand, METTS approaches are based on a {\it variational} representation of the quantum many-body states in the form of an MPS, and their remarkable accuracy relies on the extraordinary representation power of these wave functions. The fact that entanglement at finite temperatures remains under control when the initial state is a random product state has enabled some outstanding progress toward understanding the thermodynamic behavior of frustrated magnets \cite{Wietek2021}.

In this work, we take a similar route, but using neural network wave functions instead. Although our considerations are general, for illustration purposes we here focus on the simplest form, a restricted Boltzmann machine (RBM).
Same as MPS wave functions, RBMs are agnostic to the underlying physics of the problem, and hold a remarkable representation power. 

This manuscript is organized as follows: In sec.\ref{sec:METTS} we review how typicality can be used to calculate thermodynamic properties of quantum systems; sec. \ref{sec:method} discusses the practical implementation of these ideas using variational Monte Carlo; sec.\ref{sec:results} demonstrates the methods with applications to one- and two-dimensional quantum spin systems; finally, we close with a discusssion.

\section{Typical thermal states}
\label{sec:METTS}
In this section we follow the reasoning outlined in Ref. \onlinecite{White2009} to describe thermal averages in terms of typical states.
We consider a set of initial states $\{ |\phi_0(\xi) \rangle  \}_{\xi}$ with $\xi$ drawn from a probability distribution function $P_0(\xi)$ such that 
$$ \int d\xi P_0(\xi) \, | \phi_0(\xi) \rangle \langle \phi_0(\xi) | = 1.  $$

We have not explicitly introduced a particular form for these states, yet. The index $\chi$ can represent variational parameters, or just an index to label them. Each drawn initial state is evolved in imaginary time $\beta=1/T$ as:

$$ | \phi(\beta;\xi) \rangle = e^{-\frac{1}{2}\beta \hat{H}}|\phi_0(\xi) \rangle 
$$

Introducing $ Z(\beta;\xi) = \langle \phi(\beta;\xi)|\phi(\beta;\xi) \rangle
$, the partition function can expressed as  $\mathcal{Z}(\beta) = <Z(\beta,\xi)>_{P_0}$ and the evolution operator in imaginary time as 
$$ \frac{1}{\mathcal{Z}(\beta)}e^{-\beta \hat{H}} = \frac{ <|\phi(\beta;\xi)\rangle\langle\phi(\beta;\xi)|>_{P_0}}{<Z(\beta;\xi)>_{P_0}}
$$
Then, for any observable given by operator $\hat{A}$, its thermal average can be expressed as %
$$
\mathcal{A}(\beta)= \frac{<Z(\beta;\xi)A(\beta;\xi)>_{P_0}}{<Z(\beta;\xi)>_{P_0}}, 
$$
where
\[
A(\beta;\xi)=\left.\frac{\langle\phi|\hat{A}|\phi\rangle }{\langle\phi|\phi\rangle}\right|_{(\beta;\xi)}
\]
is the ``local'' expectation value of the operator in state $| \phi(\beta;\xi) \rangle$.
Introducing a finite temperature distribution $P_{\beta}(\xi) = (Z(\beta,\xi)/\mathcal{Z}(\beta))P_0(\xi)$, the thermal average can be expressed compactly as $\mathcal{A}(\beta)= <A(\beta;\xi)>_{P_{\beta}}$.  

The importance weights  $Z(\beta,\xi))/\mathcal{Z}(\beta)$ can be obtained without explicitly calculating $\langle \phi(\beta;\xi)|\phi(\beta;\xi) \rangle$ for each $\beta$.  Instead we only need the initial $Z(0;\xi)$ and, for each $\beta$, the expectation value of Hamiltonian:
$$ E(\beta;\xi)=\left.\frac{\langle\phi|\hat{H}|\phi\rangle }{\langle\phi|\phi\rangle}\right|_{(\beta;\xi)}
$$
Then, we can exploit that the imaginary time evolution gives 
$$ \frac{\partial}{\partial \beta} \ln{Z(\beta;\xi)} = - E(\beta;\xi).
$$
Hence,
$$ Z(\beta;\xi) = Z(0;\xi)\,e^{-\int_{0}^{\beta} d\beta' E(\beta';\xi)}.
$$

Until now, we have not imposed any conditions on the structure of the random initial states. In the METTS algorithm, one chooses them over a Gaussian distribution of random product states. For instance, if the quantum degree of freedom is spins $S=1/2$ on a lattice $    \mathcal{L}$ with $N$ sites, they will be given as: 
\begin{equation}
|\phi_0(\xi)\rangle = \otimes_{l\in \mathcal{L}} \left( \frac{\xi_{l\uparrow} |\uparrow\rangle_l + \xi_{l\downarrow} |\downarrow\rangle_l}{\sqrt{|\xi_{l\uparrow}|^2 + |\xi_{l\downarrow}|^2}}   \right).
\label{phi0}
\end{equation}
In this case, the label $\xi$ represents the set of $\xi \in \mathbb{C}^{ N \times 2}$ complex numbers that are distributed according to:
$$
P_0(\xi)=\left(\frac{1}{\pi}\right)^{2N} \, \exp{\left(-\sum_{l\in\mathcal{L}}  \sum_{\sigma={\uparrow,\downarrow}}  |\xi_{l\sigma}|^2 \right)}
$$

\section{Method}
\label{sec:method}

While the concepts described in the previous section offer a prescription to calculate thermodynamic properties of quantum many-body states, exact calculations can only be carried out in small systems. In order to scale the computations to large system sizes, we require to make some sacrifices: We will use a variational representation of the wave functions. For this particular task, the mathematical structure of the wave function has to be flexible enough to be able to represent any quantum state in the spectrum, and not just the ground state. In the original formulation of METTS, matrix product states are used. In our case, we generalize the method to arbitrary variational states and we focus, as illustration, on the particular case of restricted Boltzmann machines.  

In this section we review how to carry out the time evolution of a many-body state on a variational manifold, following an elegant geometrical interpretation presented in Ref.\onlinecite{Hackl2020}. The time-evolved state will describe a ``trajectory'' that will be constrained to this manifold and will deviate from the exact trajectory in the full Hilbert space. We will find that these deviations can be partially accounted by rescaling the ``projected imaginary time'' $\tau$ such that it corresponds to an actual physical inverse temperature $\beta$. 

\subsection{Variational Imaginary Time Evolution}

For each random starting state $|\phi_0(\xi)\rangle$, the time-evolved wave function $|\phi(\beta,\xi)\rangle$ is obtained by solving the first order differential equation $\frac{\partial}{\partial \beta}|\phi(\beta,\xi)\rangle = -(1/2)\hat{H}|\phi(\beta,\xi)\rangle$, with the constraint that the new wave function has to live on the same variational manifold.  This procedure results in a series of ``equations of motion'' for the parameters $\theta(\tau)$, which are completely equivalent to the well known ``stochastic reconfiguration'' (SR) method \cite{Sorella1998, Sorella2000,Sorella2005,Neuscamman2012} --also known as ``natural gradient descent''\cite{Amari1998}-- used to carry out ground state calculations. In other words, SR consists of projecting out the ground state by evolving the state in imaginary time on the variational manifold. Therefore, any variational Monte Carlo code that implements SR, already contains all the ingredients to evolve any variational state in imaginary time. We refer the reader to a pedagogical description in Ref.\cite{Glasser2018} for a detailed derivation. 
 
      \begin{figure}
            \centering
            \includegraphics[width=\linewidth]{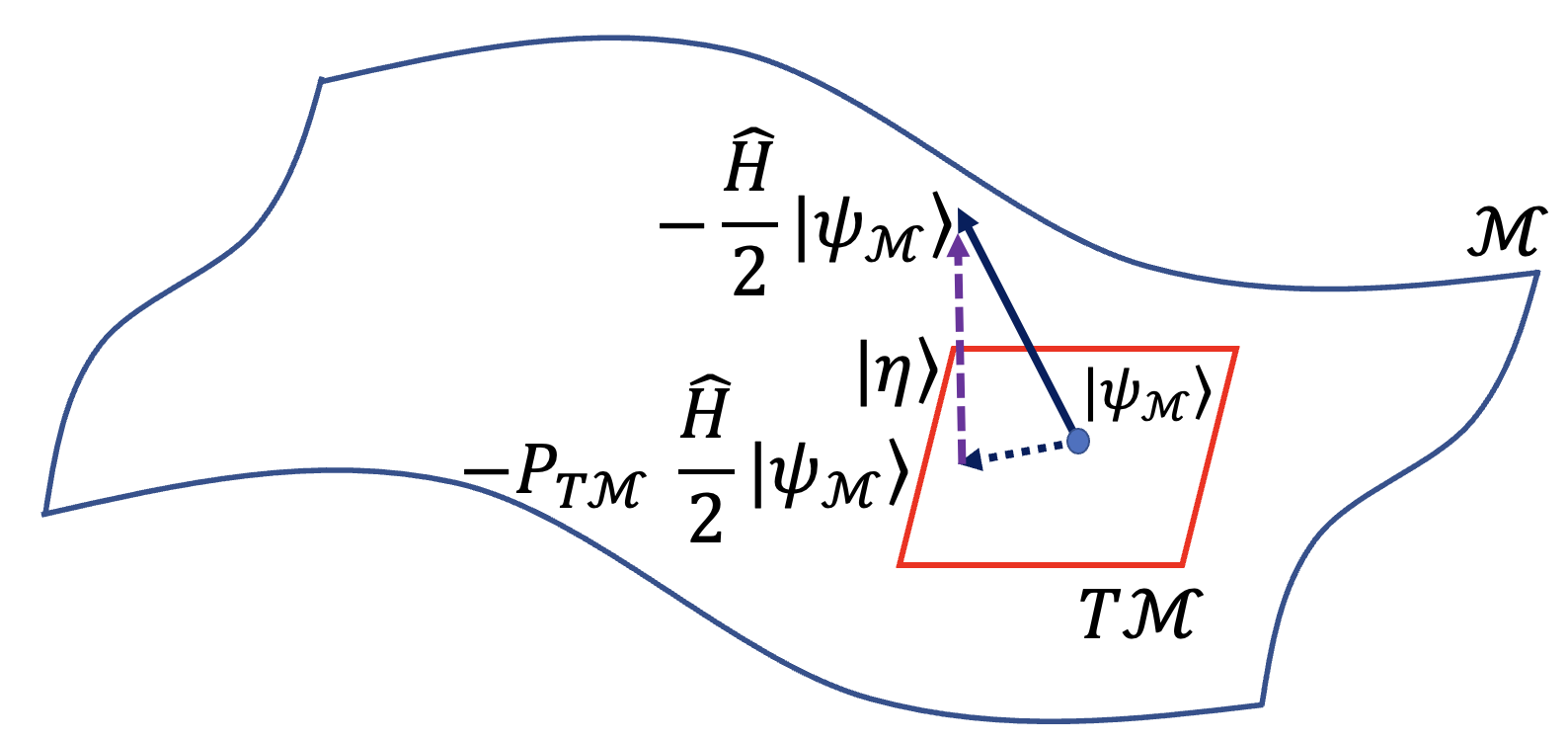}
            \caption{Cartoon illustrating imaginary time evolution in the projected variational manifold. The actual time evolved state differs from the projected one. These difference can be accounted for by a re-scaling of the temperature (see text).}
            \label{fig:fig0}
        \end{figure}

We hereby proceed to summarize projective variational imaginary time evolution for complex holomorphic variational wave functions. In this discussion we follow the notation and formalism as detailed in Ref.\onlinecite{Hackl2020}. 
The states $|\phi(\beta,\xi)\rangle$ are approximated by a class of variational wave functions $\psi_{\mathcal{M}}$ which are holomorphic in terms of parameters $\theta \in \mathbb{C}^M$ and define a sub-manifold in the Hilbert space $\mathcal{H}$,  $\mathcal{M}=\psi_{\mathcal{M}}(\mathbb{C}^M) \subset \mathcal{H}$. 

In order to carry out the imaginary time evolution within the manifold, we need to perform a local projection at each step. 
The variational parameters that best represent the time-evolved state are obtained by minimizing the projection error

\[
\left \Vert \hat{P}_{T\mathcal{H}} \left( \frac{d}{d\tau}|\psi_\mathcal{M}\rangle +\frac{1}{2}\hat{H}|\psi_\mathcal{M}\rangle \right) \right\Vert^2\]
where use the Fubini-Study metric to measure the ``distance'' between two wave functions\cite{Provost1980,Brody2001} 
with the projector $\hat{P}_{T\mathcal{H}}$ defined as
$$
\hat{P}_{T\mathcal{H}}=\left( 1-\frac{|\psi_{\mathcal{M}}\rangle\langle\psi_{\mathcal{M}}|}{\langle\psi_{\mathcal{M}}|\psi_{\mathcal{M}}\rangle}  \right). 
$$
which projects out radial dependence, given by the |$\psi_{\mathcal{M}}\rangle$ direction. 

        \begin{figure}
            \centering
            \includegraphics[width=\linewidth]{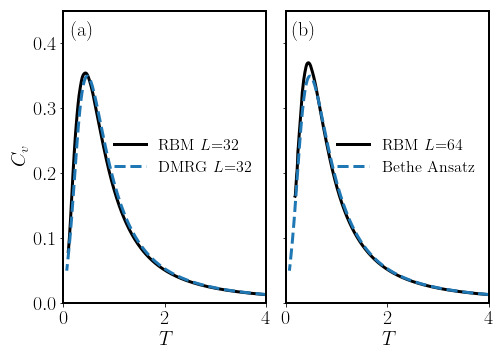}
            \caption{Specific heat of the Heisenberg chain comparing results obtained with restricted Boltzmann machines and other methods: (a) DMRG for $L=32$, (b) $L=64$ and Bethe Ansatz.}
            \label{fig:fig1}
        \end{figure}

The solution in terms of $d\theta^{\mu} / d\tau \approx  \Delta \theta^{\mu} / \Delta \tau$ is given in terms of the tangent space vectors of the variational manifold:
$$  
| v_{\mu}  \rangle =   \hat{P}_{T\mathcal{H}} \frac{\partial}{\partial \theta^{\mu}} |\psi_{\mathcal{M}}\rangle.
$$
The parameter updates $\Delta \theta$ are given in terms the system of equations:
\[
g_{\mu\nu} \Delta\theta^\nu = -\frac{\Delta\tau}{2} \frac{\langle v_\mu|\hat{H}|\psi_\mathcal{M}\rangle}{\langle\psi_\mathcal{M}|\psi_\mathcal{M}\rangle},
\]
where $g$ is the induced metric tensor:
$$
\quad g_{\mu\nu} = \frac{\langle v_{\mu}|v_{\nu}\rangle}{\langle\psi_{\mathcal{M}}|\psi_{\mathcal{M}}\rangle}.
$$

The resulting evolution of $|\psi_{\mathcal{M}}\rangle$ can be expressed compactly as
$$
\hat{P}_{T\mathcal{H}} \frac{d}{d\tau}|\psi_{\mathcal{M}}\rangle = -\frac{1}{2} \hat{P}_{T\mathcal{M}}\hat{H}|\psi_{\mathcal{M}}\rangle,
$$
where we introduce the variational tangent space projector:
$$
\hat{P}_{T\mathcal{M}} = G^{\mu\nu}\frac{| v_\mu \rangle\langle v_\nu |}{\langle\psi_\mathcal{M}|\psi_\mathcal{M}\rangle}; \quad G^{\mu\nu}g_{\nu\sigma}=\delta^{\mu}_{\,\sigma}
$$



 As discussed above, each initial trial wave function is drawn from a distribution $P_0(\xi)$ and the parameters are evolved in imaginary time by a small fixed time step $\Delta\tau$. At each time step, expectation values of the energy, variance of the energy, and observables of interest are calculated:
\begin{eqnarray} 
{E}(\theta(\tau;\xi))&=&\left.\frac{\langle\psi|\hat{H}|\psi\rangle }{\langle\psi|\psi\rangle}\right|_{\theta(\tau;\xi)}, \\
 {\sigma}^2(\theta(\tau;\xi)) &=&\left.\frac{\langle\psi|(\hat{H}-{E})^2|\psi\rangle }{\langle\psi|\psi\rangle}\right|_{\theta(\tau;\xi)},\\
 {A}(\theta(\tau;\xi))&=&\left.\frac{\langle\psi|\hat{A}|\psi\rangle }{\langle\psi|\psi\rangle}\right|_{\theta(\tau;\xi)}.
\end{eqnarray}
These expectation values are then averaged over many initial realizations of $\xi$.

       \begin{figure}
            \centering
            \includegraphics[width=\linewidth]{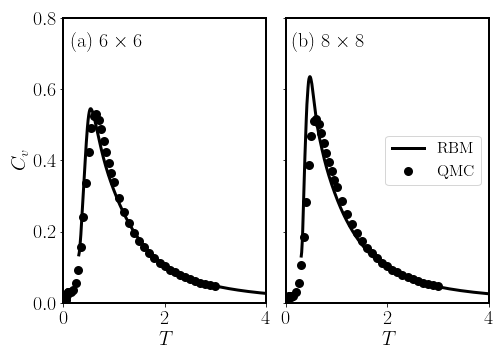}
            \caption{Specific heat of the 2D Heisenberg model comparing results obtained with restricted Boltzmann machines and quantum Monte Carlo for two system sizes (a) $6\times6$ and (b) $8\times8$.}
            \label{fig:fig2}
        \end{figure}

\subsection{Beta Correction}

Since the approximate imaginary time evolution corresponds to projecting onto the variational manifold, the time-evolved state will deviate from the exact one. This deviation from the true path can be decomposed locally into two contributions (see Fig.\ref{fig:fig0}):
\begin{equation}
\hat{P}_{T \mathcal{H}}\frac{d}{d\tau}|\psi_\mathcal{M}\rangle =-\frac{1}{2}\gamma\hat{P}_{T \mathcal{M}}\hat{H}|\psi_\mathcal{M}\rangle +|\eta\rangle
\end{equation}
where $|\eta\rangle$ is some error direction orthogonal to both, $|\psi_\mathcal{M}\rangle$ and $(\hat{H}-{E})|\psi_\mathcal{M}\rangle  $. The factor $ \gamma\in \left[ 0,1 \right] $ represents the fraction of the distance traveled in the exact imaginary time direction, that can be explicitly calculated as  
\begin{equation}
\gamma = \frac{\langle\psi|(\hat{H}-{E})  \hat{P}_{T \mathcal{M}}(\hat{H}-{E})|\psi\rangle}{\langle\psi|(\hat{H}-{E})^2|\psi\rangle}= -\frac{1}{{\sigma}^2}\frac{d{E}}{d\tau}
\label{eq:beta}
\end{equation}

Here we assume that the contribution from the error direction $|\eta\rangle$ is fairly negligible (which is a reasonable assumption for imaginary time evolution as opposed to real time evolution).  Thus, the remaining error corresponds to the parametrization of the imaginary time. Therefore, we re-parameterize $\tau$ in terms of $\beta$ as:
$$
\frac{d\beta}{d\tau}=\gamma;  \, \beta(\tau;\xi)=-\int_{0}^{\tau}d\tau' \, \left. \frac{1}{{\sigma}^2}\frac{d{E}}{d\tau'}\right|_{\theta(\tau';\xi)}
$$
Following Eq.(\ref{eq:beta}), this parameterization also enforces that $dE / d\beta = - \sigma^2$.

\subsection{Restricted Boltzmann Machines}

We have not made any assumptions about the variational form of the wave functions thus far. In order to demonstrate the application of these ideas, we focus on a particular example, a restricted Boltzmann machine(RBM).
An RBM wave function for a system of $N$ spins $S=1/2$ ($s={\uparrow,\downarrow}$) and $M$ hidden variables is defined as:
$$
|\psi(\theta)\rangle = \sum_{\{\vec{s}\}} \psi(\vec{s};\theta) |\vec{s}\rangle
$$
where $\vec{s}=(s_1,s_2,\cdots,s_N)$ and
$$
\psi(\vec{s};\theta) = e^{\sum_{l\in\mathcal{L}} a_ls_l}\prod_{i=1}^{M} \cosh\left(   b_i + \sum_{l\in\mathcal{L}} W_{il} \cdot s_l \right)
$$
It is parametrized by a set of complex values $\theta = (a,b,W)\in\ \mathbb{C}^N \times \mathbb{C}^M \times \mathbb{C}^{M \times N}$ that are used as variational parameters to minimize some cost function. This cost function is usually a measure of the ``distance'' between $|\psi(\theta)\rangle$ and a target wave function.   
            \begin{figure}
            \centering
            \includegraphics[width=\linewidth]{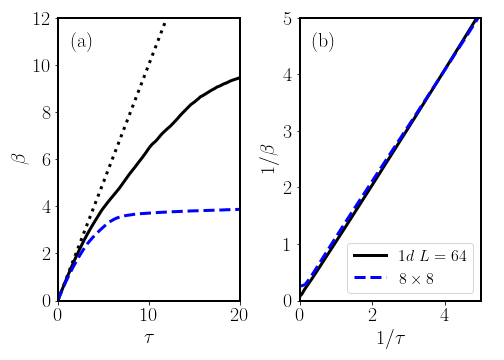}
            \caption{(a) Actual inverse temperature $\beta$ versus 
             projected imaginary time $\tau$. We show results for both 1d and 2d lattices. The dotted line corresponds to a 1:1 perfect scaling. The differences arise from projecting on the variational manifold (see text). (b) Same data as panel (a), but showing the physical temperature $T=1/\beta$.
             }
            \label{fig:fig3}
        \end{figure}
           
Besides being proposed as ground state estimators for variational calculations \cite{Carleo2017,Saito2017,Cai2018,Glasser2018}, their representation power has been instrumental to a number of other applications, such as the calculation of spectral functions \cite{Hendry2019, Hendry2021, Koch2021}. 

\subsection{RBM Initialization}
In order to implement the procedure with RBMs, it is necessary to first draw a random set of initial product wave functions as described in sec.\ref{sec:METTS}. Fortunately, it is possible to exactly represent (up to an overall constant) a state $|\phi_0(\xi)\rangle$ in Eq.(\ref{phi0}) by setting $b,W=0$ and matching the $a_l$ bias term to the corresponding random spin at $l$
$$
a_l = \frac{1}{2}\ln( \xi_{l+} / \xi_{l-} )
$$
In practice, a very small Gaussian noise has to be added to $W,b$ such that the derivatives needed in the imaginary time evolution are not zero. Notice that there is no requirement that the initial state has to be a product state, so the validity of the method is not affected.

\section{Numerical results}
\label{sec:results}

We have implemented the imaginary time evolution for the spin $S=1/2$ Heisenberg model in one and two dimensions:
\[
\hat{H}=\sum_{\langle i,j\rangle}\vec{S}_i\cdot\vec{S}_j
\]
where the sum sums over nearest neighboring sites on a one-dimensional chain, or a square lattice.

Our results were obtained by averaging over 50 initial random states using the same number of hidden and visible variables. In Fig.\ref{fig:fig1} we show the specific heat for a chain with $L=32$ and $L=64$ sites and periodic boundary conditions, together with finite-temperature density matrix renormalization group  \cite{Feiguin2005a} and Bethe ansatz\cite{Klumper2000} results for comparison. The curves are barely distinguishable in this scale. Results for two-dimensional systems (Fig.\ref{fig:fig2}) are less accurate, as compared to quantum Monte Carlo data for $6\times 6$ and $8 \times 8$ lattices \cite{Sandvik1999,Hoglund2004,Sandvik_data}. Besides a sharper peak in the specific heat for $8\times8$, there seems to be an apparent horizontal shift. This can be explained in terms of the beta correction. In Fig.\ref{fig:fig3} we show the actual $\beta$ as a function of the projected imaginary time $\tau$ for two typical runs. While the curves seem to follow an apparent 1:1 scaling at small $\beta$, deviations appear as we approach zero temperature. The reasons why these deviations increase could be multiple, but we believe the most significant ones can be due to: (i) the presence of a gap in the spectrum or (ii) a small overlap with the actual ground state due to the random initial directions of the spins, or (iii) a poor variational wave function. Still, the accuracy can be improved by increasing the number of hidden variables/variational parameters in the RBM wave functions. The most promising route to improve the representation power of RBMs is by introducing lattice symmetries\cite{Nomura2021} which we have not attempted here. 

\section{Conclusions}
\label{sec:conclusions}

We have described a method to carry out thermodynamic simulations of quantum many-body models using typicality and variational representations of quantum states. We generalize the idea of METTS to arbitrary variational forms and efficiently carry out the imaginary time evolution using natural gradient descent (or stochastic reconfiguration). The underlying mathematical structure of the wave functions plays a crucial role in terms of the accuracy of the method. In particular, the wave functions have to be able to represent any state along the imaginary time path. In this work, we pick the particular form of restricted Boltzmann machines as a proof of concept illustration due to their versatility and representation power. We show that the path they follow in the variational manifold differs slightly from the actual imaginary time evolution. We have found that these deviations can be accounted for by correcting the temperature with a rescaling factor that can be easily and systematically calculated at every time step. The method does not suffer from the sign problem and offers an alternative to matrix product states for studying two dimensional models with frustration.

\acknowledgements
The authors are grateful to the National Science Foundation for support under grant No DMR-1807814. We thank Anders Sandvik for generously sharing his QMC data with us.


%

\end{document}